\documentclass[epj]{svjour}
\usepackage{graphics}

\begin{document}

\title{Backward pion-nucleon scattering}
\author{F.~Huang\inst{1}, A.~Sibirtsev\inst{2,3}, J.~Haidenbauer\inst{4,5},
S.~Krewald\inst{4,5} and U.-G.~Mei{\ss}ner\inst{2,4,5}
}                     


\institute{Department of Physics and Astronomy, The University of
Georgia, Athens, Georgia 30602, USA \and
Helmholtz-Institut f\"ur Strahlen- und Kernphysik (Theorie)
und Bethe Center for Theoretical Physics,
Universit\"at Bonn, D-53115 Bonn, Germany  \and
Excited Baryon Analysis Center (EBAC), Thomas Jefferson National
Accelerator
Facility, Newport News, Virginia 23606, USA
\and Institut f\"ur Kernphysik and J\"ulich Center for Hadron Physics,
Forschungszentrum J\"ulich, D-52425 J\"ulich, Germany
\and Institute of Advanced Simulation,
Forschungszentrum J\"ulich, D-52425 J\"ulich, Germany
}

\date{Received: date / Revised version: date}

\abstract{A global analysis of the world data on 
differential cross sections and polarization asymmetries
of backward pion-nucleon scattering 
for invariant collision energies above 3 GeV is performed in a Regge model.
Including the $N_\alpha$, $N_\gamma$, $\Delta_\delta$ and
$\Delta_\beta$ trajectories, we reproduce both angular distributions and 
polarization data for 
small values of the Mandelstam variable $u$, in contrast to previous analyses.
The model amplitude is  used to obtain evidence for baryon resonances
with mass below 3~GeV. Our analysis suggests a $G_{39}$ resonance with a 
mass of 2.83~GeV as member of the $\Delta_{\beta}$ trajectory from the
corresponding Chew-Frautschi plot.
} 

\authorrunning{F. Huang {\it et al.}}

\maketitle

\section{Introduction}

The excitation spectrum of the nucleon in the energy range up to 2.5 GeV is
presently investigated with electromagnetically induced reactions at JLab,
ELSA and Mainz. A major progress is expected from the experimental
determination of polarization observables and exclusive measurements. New
theoretical methods appear to be necessary for the analysis of the incoming
data, as traditional partial wave analyses have to deal with a large number
of parameters.  For instance, the methods developed at EBAC at JLab  are
based on theories of non-resonant meson dynamics which, taken together with
the resonant contributions,  describe a variety of reactions. However at
large energies, two-body reactions show regular features which with
very high likelihood 
are not due to individual resonances. For energies above 2~GeV,
the angular distributions are strongly forward peaked and show a smooth
energy dependence. Moreover, at backward angles there is another
regularity. In that angular region many reactions show a rise of the cross
sections, with a magnitude that depends smoothly on the energy. 

Most of our knowledge about resonances with large spin has been obtained
from the study of pion scattering near 180$^0$. A summary of models for
high energy meson-nucleon backward scattering is given in 
refs.~\cite{Barger67,Barger68,Berger71,Gregorich71,Storrow75}. These models are
based on Regge phenomenology using parameters  determined by a systematic
analysis of the backward differential cross section at different energies
that were available before 1972. None of these models can describe
simultaneously the data on differential cross sections and polarizations
available at different energies.  Experimental results at certain fixed
energies were reproduced by introducing non-Regge terms \cite{Birsa77} in
addition to the standard Regge amplitudes. There are data for $\pi N$
backward scattering published after 1974
\cite{Baglin75,Jacholkowski77,Ghidini82,Baker83,Armstrong87}, but
unfortunately, most of these data have never been analyzed in the framework
of the previous Regge models.

A major goal of this work is to analyze differential cross sections and
polarization data of $\pi N$ backward scattering for invariant collision
energy $\sqrt{s} \geq $ 3 GeV. We include experimental results available
for the reactions $\pi^+p \to \pi^+p$, $\pi^-p \to \pi^-p$ and  $\pi^-p \to \pi^0n$. 
Regge phenomenology is applied in order to fix the reaction
amplitude given by the contribution from four exchange trajectories, namely
$N_\alpha$, $N_\gamma$, $\Delta_\delta$ and $\Delta_\beta$. These
trajectories are parameterized by real linear functions of the 
squared four-momentum transfer $u$. Using $505$ data points, we obtain a 
$\chi^2$ per data point of $1.84$. 

Let us briefly recall the status of the published Regge phenomenology for
backward pion-nucleon scattering. The differential cross section data for
the reactions $\pi^+p \to \pi^+p$, $\pi^-p \to \pi^-p$, and $\pi^-p \to
\pi^0n$  can be described by  three baryon trajectories, called $N_\alpha$,
$N_\gamma$ and $\Delta_\delta$, which start with the nucleon, the
$D_{13}(1520)$, and the $\Delta_{33}(1232)$, respectively. The $N_\alpha$
and $\Delta_\delta$ trajectories are the leading baryon trajectories for
the $u$-channel isospin $I_u=1/2$ and $I_u=3/2$ reactions. The dip
structure at $u \approx -$0.15~GeV$^2$ in the backward $\pi^+p \to \pi^+p$
differential cross section is due to a zero in the $N_\alpha$ trajectory. A
third trajectory $N_\gamma$ is needed because of the differences in the dip
structure of the $\pi^+p \to \pi^+p$ and $\pi^-p \to \pi^0n$ differential
cross sections. First polarization data were obtained after 1971
\cite{Aoi71,Dick72,Dick73,Birsa76}. None of the Regge models available at
that time could describe those data. Furthermore, for the $\pi^-p \to
\pi^-p$ reaction the $N_\alpha$ and $N_\gamma$ exchanges do not contribute.
Thus, within the frame of pole-exchange Regge models there is no relative
phase between the spin flip and spin non-flip amplitudes from the
$\Delta_\delta$ trajectory alone. As a consequence, the available models
predicted zero polarization for the $\pi^-p \to \pi^-$p reaction, in
contradiction to the polarization data.

Several attempts were made in order to resolve the conflict between
polarization data and the models. First, let us mention the Regge model of ref.
\cite{Gregorich71} which originally included only the $N_\alpha$ and $\Delta_\delta$
trajectories. The polarization predicted by this model for the $\pi^-p \to
\pi^-p$ reaction was not zero because the trajectories were considered as
complex functions parameterized in a quite sophisticated way. Later on it
was found that the predictions of this model disagreed with the
polarization data and the model was substantially modified \cite{Park74}
through additional inclusion of the $N_\gamma$ trajectory and by using
sophisticated parametrizations of the vertex functions. After that the
differential cross section and polarization at two beam momenta, $5.91$ and
6~GeV, were well fitted. It was unclear whether this modified model still
is able to reproduce data on differential cross sections available at other
energies that were analyzed originally \cite{Gregorich71}, because the
systematic analysis was not repeated. 

In the Regge analysis of ref. \cite{Mir81} the $N_\alpha$ and
$N_\gamma$ trajectories were taken as real functions. But,
following refs. \cite{Storrow75,Storrow73} the $\Delta_\delta$
contribution was parameterized differently with regard to the real and
imaginary part of the amplitude. That allows to obtain relative phases
between the spin flip and non-flip amplitudes as required by the
non-zero polarization for the $\pi^-p \to \pi^-p$ process. It was
shown that this model reproduces well the data on differential cross
sections at pion momenta above 23 GeV and the $\pi^+p \to \pi^+p$
polarization \cite{Aoi71}. However, these calculations could not
describe the polarization data \cite{Dick73} in the $\pi^-p \to \pi^-p$
reaction. Although a reasonable description of the data was achieved, except
for the $\pi^-p$ polarization, one should note that the 
modification of the $\Delta_\delta$ amplitude employed is not conventional
in Regge phenomenology. Furthermore, it was not shown whether this
model can reproduce the data at momenta below 23~GeV.

Another effort to describe the data on differential cross sections and
polarizations was presented in ref. \cite{Birsa77}. Only data at the pion
momentum of 3.5~GeV
\cite{Birsa76,Banaigs68,Banaigs69,Bradamante73,DeMarzo75} and 6~GeV
\cite{Dick72,Dick73,Owen69,Boright70} were included in the analysis
performed with two different Regge models. The first one takes into account
the $N_\alpha$, $N_\gamma$ and the modified \cite{Storrow75,Storrow73}
$\Delta_\delta$ contributions discussed above. The second model
\cite{Donnachie74} accounts for amplitudes given by the standard
$N_\alpha$, $N_\gamma$ and $\Delta_\delta$ Regge pole terms, but includes
in addition a coherent background amplitude which is ascribed to quark
rearrangement processes \cite{Gunion72}. It was shown that both models
reproduce the data well when fitted separately to the experimental results
at momenta of 3.5 GeV and 6 GeV. But it was not possible to obtain a
reasonable description of the experiments within a simultaneous fit of the
3.5 GeV and 6 GeV data. Therefore, it was speculated that the polarization
data for the $\pi^-p \to \pi^-p$ reaction might indicate that some of the
amplitudes have an energy dependence that differs from the power law in
invariant collision energy, which is typical for Regge phenomenology.
However, any definite conclusion requires more systematic and comprehensive
theoretical studies of backward pion scattering, which was not done.

The present study shows that the addition of a second Delta trajectory
leads to an economic description of backward pion-nucleon scattering for
large energies in terms of a simple Regge phenomenology. The amplitudes
obtained at high energies allow an extrapolation to lower energies. We
inspect how rapidly the Regge phenomenology starts to deviate from the data
at invariant collision energies $2.4 \le \sqrt{s} \le 3$ GeV. The analysis
concentrates on available polarization data that are expected to be
sensitive to  possible contributions of  high mass resonances. The
indicated energy range is chosen for two principal reasons. First, the
energy dependence of the differential cross section for the pion-nucleon
scattering at 180$^0$ shows some structures at these energies. Second, many
earlier analyses \cite{Hoehler83,Hendry78,Cutkosky79,Koch80} found evidence
for excited baryons with masses within this energy range. We study in
detail the data available for $\pi^+p \to \pi^+p$, $\pi^-p \to \pi^-p$ and
$\pi^-p \to \pi^0n$ scattering at 180$^0$ and calculate the confidence
level for the discrepancy between the data and our results. That allows us
to estimate whether the oscillations of differential cross section around
the continuation of the Regge result is of systematic or rather of
statistical nature.

Finally we investigate the relation between the baryon trajectories
fixed by our analysis in the scattering \, region and the \, baryon
spectrum.
Indeed the \, ordering of the hadronic states according to the Regge
trajectories in
the Chew-Frautschi plot is one of the most remarkable features of
the Regge phenomenology. However, the Regge classification of baryons
in many studies \cite{Storrow72,Mir83,Glozman02,Goity07} was done by
using known or predicted states and not scattering data. Here we
address the question whether the trajectories obtained in the
analysis of backward pion scattering are the same as those given by
the known baryon spectrum.

The paper is organized as follows. In sect. 2, we introduce our formalism. 
Sect. 3 provides a comparison of the results of our calculations with data
on differential cross sections and polarizations at invariant collision
energies above 3 GeV. In sect. 4 the amplitude is extrapolated to lower
energies. The paper ends with a Summary. An appendix summarizes the available
world data set on backward pion-nucleon scattering.

\vfill

\pagebreak

\section{Formalism}
The differential cross section for backward pion-nucleon scattering reads
\begin{eqnarray}
\frac{d\sigma}{du}=\frac{|{\cal M}^{++}|^2+|{\cal M}^{+-}|^2}{64\pi
sq^2}.
\end{eqnarray}
Here, the $s$-channel helicity  non-flip and flip amplitudes are called ${\cal M}^{++}$ and
${\cal M}^{+-}$, respectively,  
while $q$ denotes the pion momentum in the $s$-channel center-of-mass
(CM) system, using the notations
s,t, and u for the Mandelstam variables.
The polarization asymmetry is given by
\begin{eqnarray}
P=\frac{2\,{\rm Im}\!\left[{\cal M}^{++}{{\cal
M}^{+-}}^\ast\right]}{{|{\cal M}^{++}|}^2 + {|{\cal M}^{+-}|}^2}.
\end{eqnarray}
Taking the amplitudes with specified $u$-channel isospin, we can
write the $\pi N$ backward scattering amplitudes as
\begin{eqnarray}
{\cal M}^{\pi^+p\,\to\, \pi^+ p}&=&\frac{2}{3}{\cal
M}^N+\frac{1}{3}{\cal M}^\Delta, \\
{\cal M}^{\pi^-p\,\to\, \pi^- p}&=&{\cal
M}^\Delta, \\
{\cal M}^{\pi^-p\,\to\, \pi^0 n}&=&\frac{\sqrt{2}}{3}{\cal
M}^N-\frac{\sqrt{2}}{3}{\cal M}^\Delta,
\end{eqnarray}
where the $N$ and $\Delta$ superscripts denote $u$-channel
isospin $1/2$ and $3/2$ contributions, respectively.
The $s$-channel helicity amplitudes are expressed in terms of the
invariant amplitudes $A$ and $B$ as
\begin{eqnarray}
{\cal
M}^{++}&=&2\left[m_NA+\left(E_N\sqrt{s}-m_N^2\right)B\right]{\rm
cos}(\theta/2),  \\
{\cal M}^{+-}&=&2\left[E_NA+m_N\left(\sqrt{s}-E_N\right)B\right]{\rm
sin}(\theta/2),
\label{flip}
\end{eqnarray}
where $E_N$ refers to the energy of the nucleon in the $s$-channel 
CM system, and $\theta$ is the $s$-channel scattering angle.

We parameterize the invariant amplitudes $A$ and $B$  \cite{Hoehler83} by
\begin{eqnarray}
A &=& \sum_i \beta_i^A(u) \frac{
\zeta_i(u)}{\Gamma\!\left(\alpha_i-1/2\right)}
\left(\frac{s}{s_0}\right)^{\alpha_i-1/2}, \\
B &=& \sum_i \beta_i^B(u) \frac{
\zeta_i(u)}{\Gamma\!\left(\alpha_i-1/2\right)}
\left(\frac{s}{s_0}\right)^{\alpha_i-1/2}, 
\label{amplitude}
\end{eqnarray}
where $s_0 = $ 1 GeV$^2$ is a scaling factor
and $\zeta_i(u)$ is the Regge propagator,
\begin{eqnarray}
\zeta_i(u)= \frac{1+{\cal S}_i \, {\rm
exp}\!\left[-i\pi\!\left(\alpha_i(u)-1/2\right)\right]}{{\rm sin}
[\pi\!\left(\alpha_i(u)-1/2\right)]},
\label{signature}
\end{eqnarray}
with ${\cal S}_i$ denoting the signature of the trajectory. The $i$-th baryon
Regge trajectory, $\alpha_i$, is parameterized as a linear function of $u$,
\begin{eqnarray}
\alpha_i(u) = {\alpha_0}_i + \alpha\prime \,u~,
\end{eqnarray}
where $i$ labels the trajectories $N_\alpha, N_\gamma, \Delta_\delta$ and
$\Delta_\beta$, respectively. The slope $\alpha\prime$ and the intercept $\alpha_0$ are
determined by a fit to the data. As will be discussed later,
we take the same slope parameter for all four trajectories.

\begin{table}[t]
\begin{center}
\caption{\label{notation} The leading baryon trajectories.
The last column shows the parity partners that have the same signature 
but opposite parity.}
\renewcommand{\arraystretch}{1.2}
\begin{tabular}{|c|c|c|c|c|}
\hline
  Trajectory     & Isospin & Parity & Signature & Partner \\
\hline
 $N_\alpha$      &  $1/2$  &  $+$   & $+$ &  $N_\beta$ \\
 $N_\gamma$      &  $1/2$  &  $-$   & $-$ &  $N_\delta$ \\
 $\Delta_\delta$ &  $3/2$  &  $+$   & $-$ &  $\Delta_\gamma$  \\
 $\Delta_\beta$  &  $3/2$  &  $-$   & $+$ &  $\Delta_\alpha$  \\
\hline
\end{tabular}
\end{center}
\end{table}

The baryon
trajectories used in the present work are given in table \ref{notation}. 
The signature of a trajectory is defined as
${\cal S} =  (-1)^{J-1/2}$, 
where $J$ is the baryon spin. The classification of the Regge
trajectories is given in terms of the signature ${\cal S} = {\pm}$ 1
and the parity $P = {\pm}$ 1. Thus one should consider four trajectories
for the nucleon as well as for the Delta-isobar states.  Furthermore, the
parity partners of the discussed trajectories are defined
\cite{MacDowell,Gribov} in the Regge formalism as trajectories with the
same signature but opposite parity. They are also indicated in
table \ref{notation}. The residue functions $\beta^A(u)$
and $\beta^B(u)$ for each trajectory are parameterized by
\begin{eqnarray}
\beta^A(u)&=&a+b\,u~,  \\
\beta^B(u)&=&c+d\,u~.
\end{eqnarray}

\section{Results for high energies}

\begin{figure}[b]
\centering
\resizebox{0.4\textwidth}{!}{\includegraphics{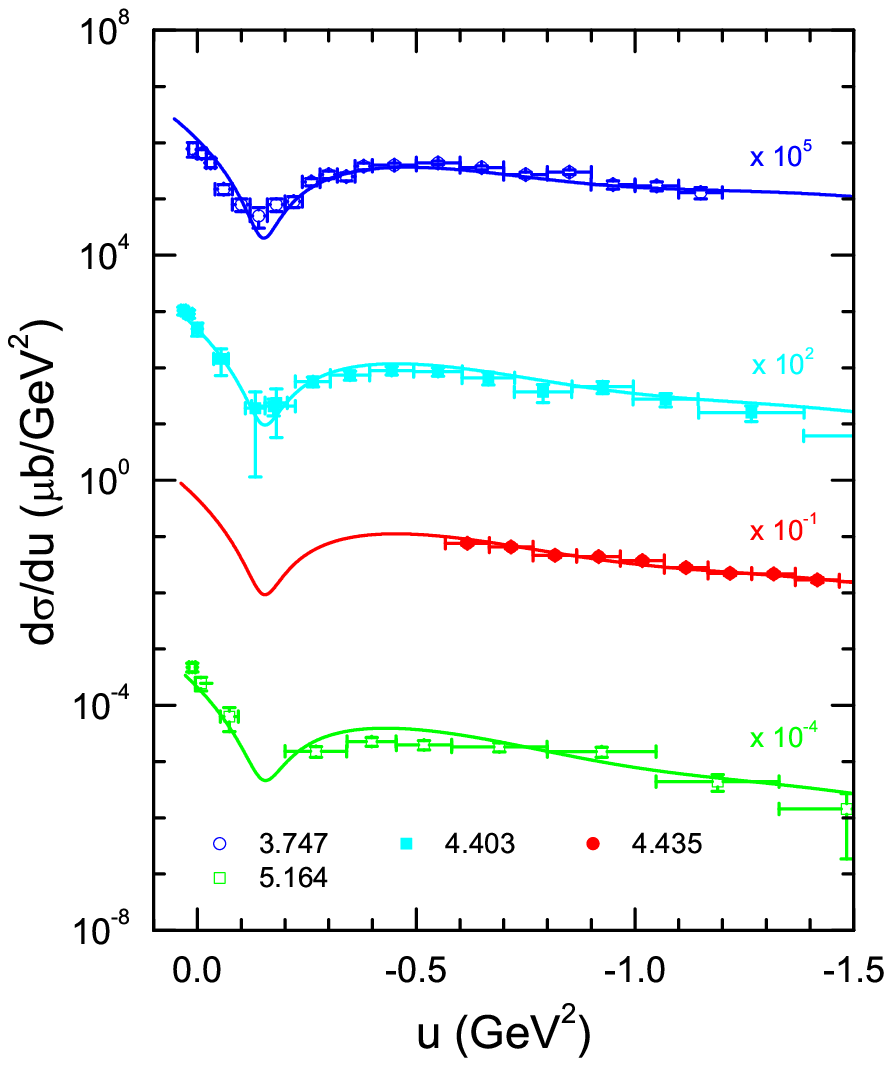}}
\caption{ \label{dsdu_pi+p_c} The differential cross section for
$\pi^+p \to \pi^+p$ backward scattering as a function of the $u$-channel
four-momentum transfer squared shown for different invariant
collision energies, $\sqrt{s}$, indicated in the legend. The references to the
data are given in tab. \ref{pi+p}. The solid lines are the results
of our model calculation.
Both data and calculations were scaled by the indicated factors.}
\end{figure}

\begin{figure}[t]
\centering
\resizebox{0.4\textwidth}{!}{\includegraphics{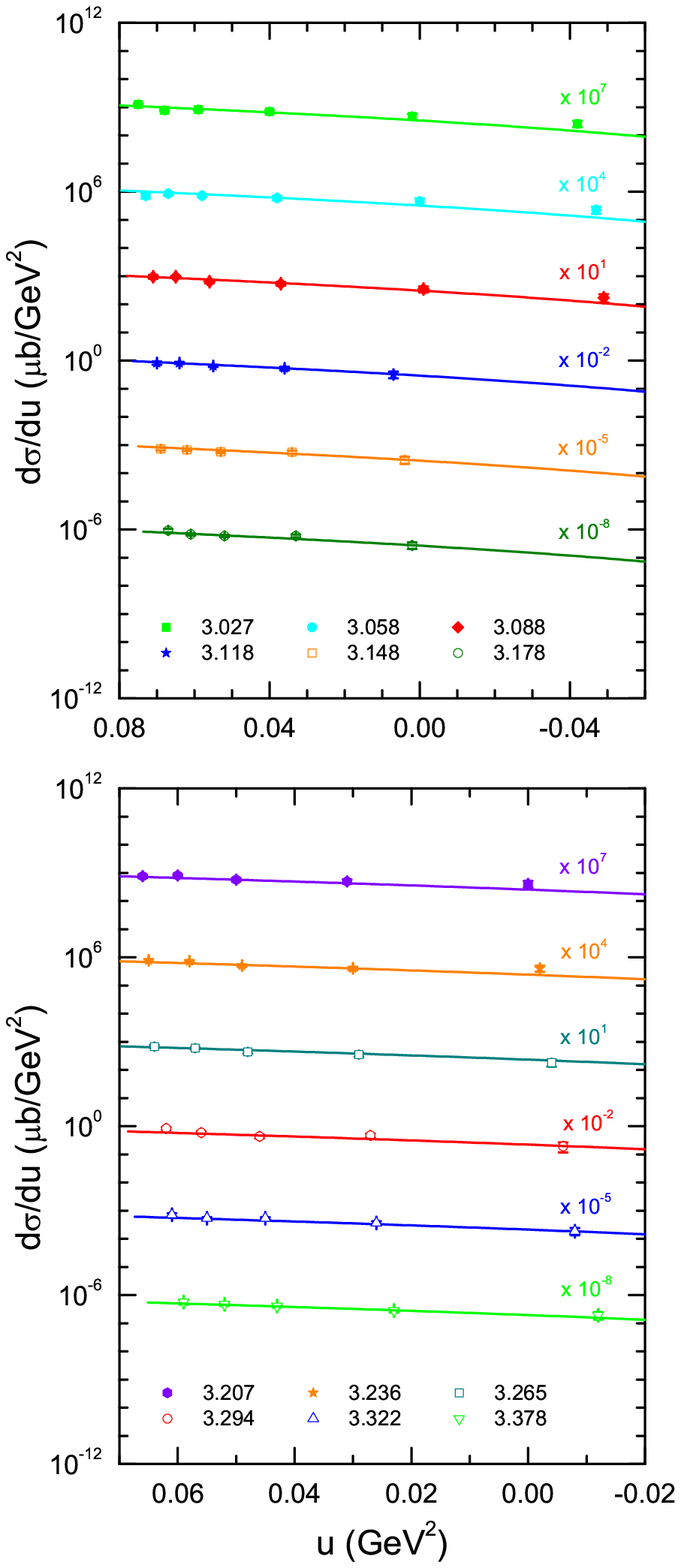}}
\caption{ \label{dsdu_pi+p_a} The differential cross section for
$\pi^+p \to \pi^+p$ backward scattering as a function of the $u$-channel
four-momentum transfer squared shown for different invariant
collision energies, $\sqrt{s}$, indicated in the legend. The
references to the data are given in tab. \ref{pi+p}. The solid lines
are the results of our model calculation. Both data and calculations were scaled by the
indicated factors.}
\end{figure}

\begin{figure}[t]
\centering
\resizebox{0.4\textwidth}{!}{\includegraphics{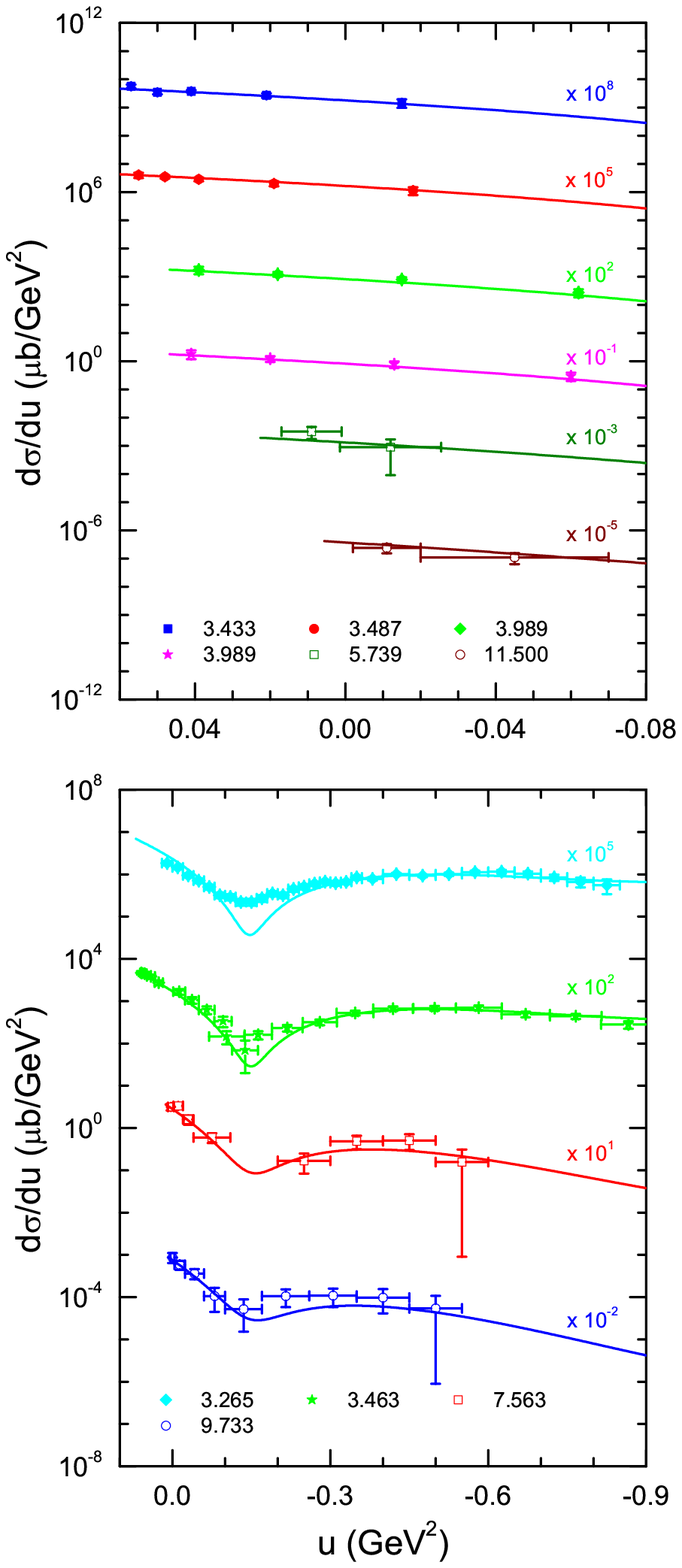}}
\caption{ \label{dsdu_pi+p_b} The differential cross section for
$\pi^+p \to \pi^+p$ backward scattering as a function of the $u$-channel
four-momentum transfer squared shown for different invariant
collision energies indicated in the legend. The references to the
data are given in tab. \ref{pi+p}. The solid lines show the results
of our model calculation.
Both data and calculations were scaled by the indicated factors.}
\end{figure}

\begin{figure}[t]
\centering
\resizebox{0.39\textwidth}{!}{\includegraphics{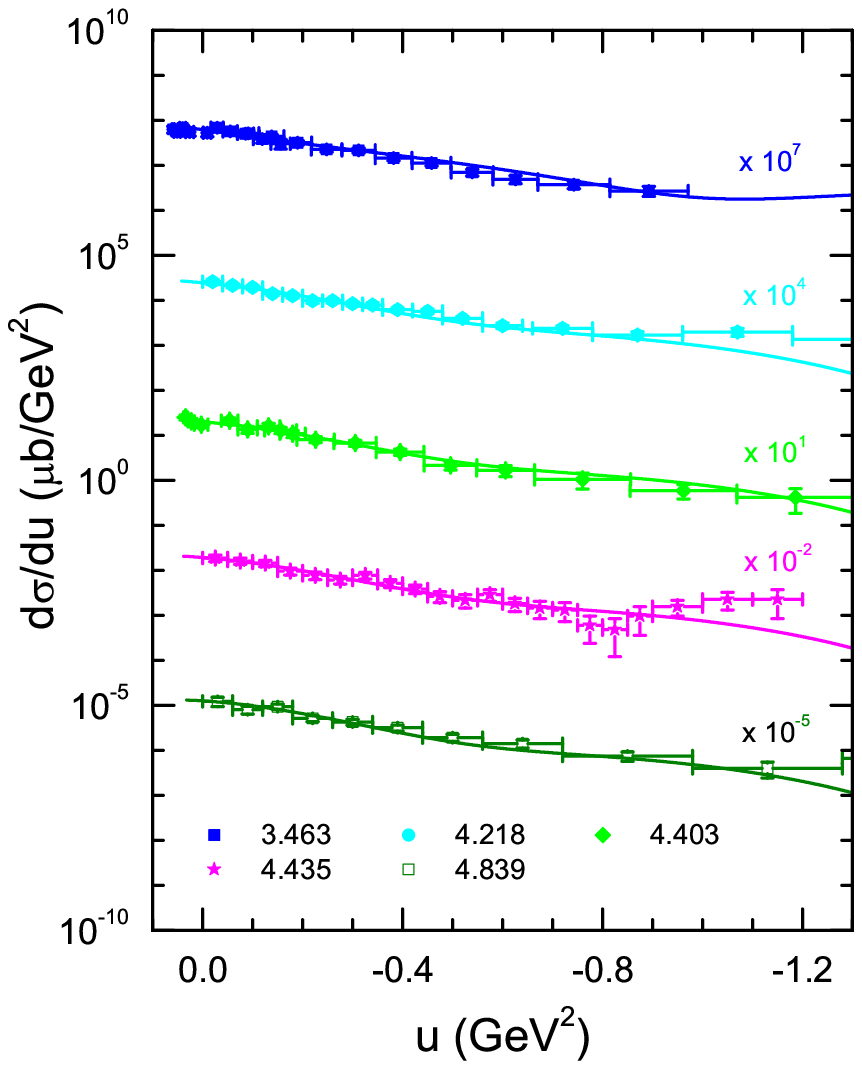}}
\caption{ \label{dsdu_pi-p_a} The differential cross section for
$\pi^-p \to \pi^-p$ backward scattering as a function of the $u$-channel
four-momentum transfer squared shown for different invariant
collision energies indicated in the legend. The references to the
data are given in tab. \ref{pi-p}. The solid lines are the results
of our model calculation.
Both data and calculations were scaled by the indicated factors.}
\end{figure}

\begin{figure}[t]
\centering
\resizebox{0.4\textwidth}{!}{\includegraphics{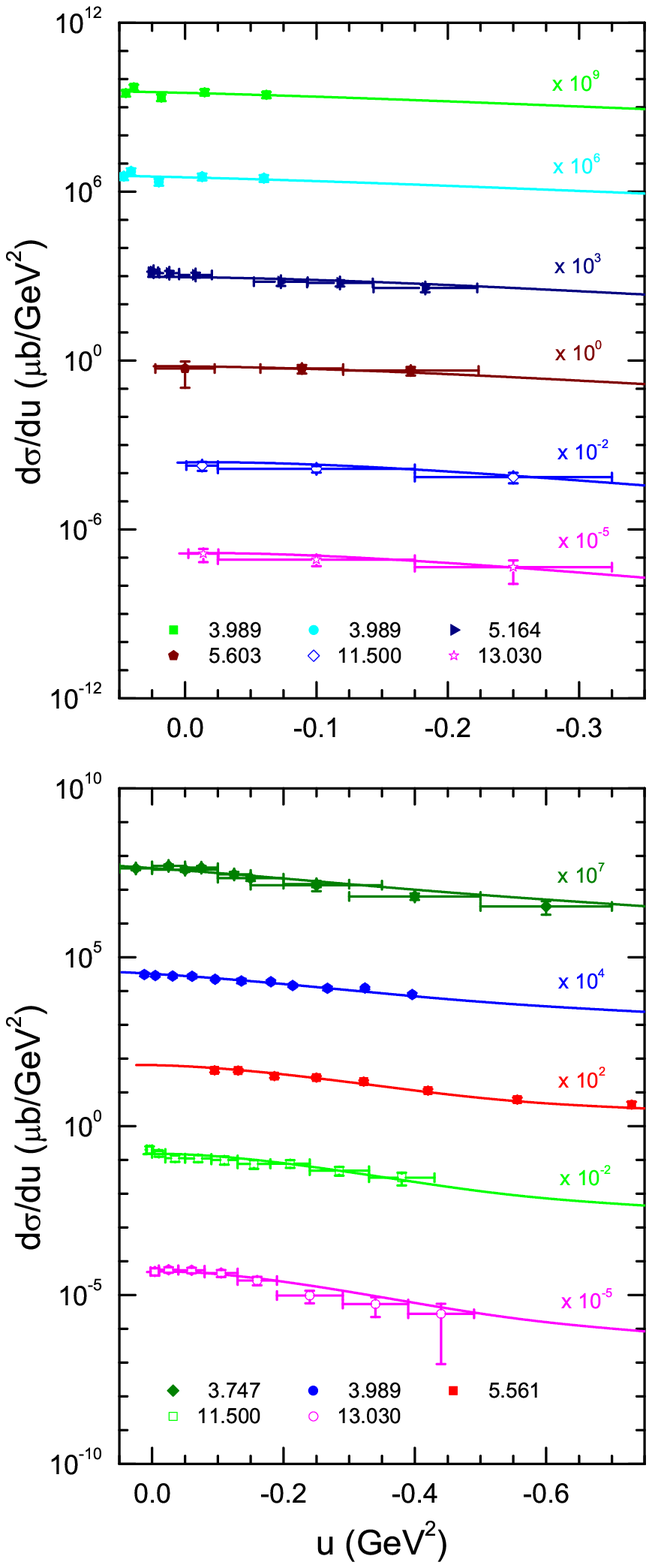}}
\caption{ \label{dsdu_pi-p_b} The differential cross section for
$\pi^-p \to \pi^-p$ backward scattering as a function of the $u$-channel
four-momentum transfer squared shown for different invariant
collision energies indicated in the legend. The references to the
data are given in tab. \ref{pi-p}. The solid lines are the results
of our model calculation.
Both data and calculations were scaled by the indicated factors.}
\end{figure}

\begin{figure}[t]
\centering
\resizebox{0.4\textwidth}{!}{\includegraphics{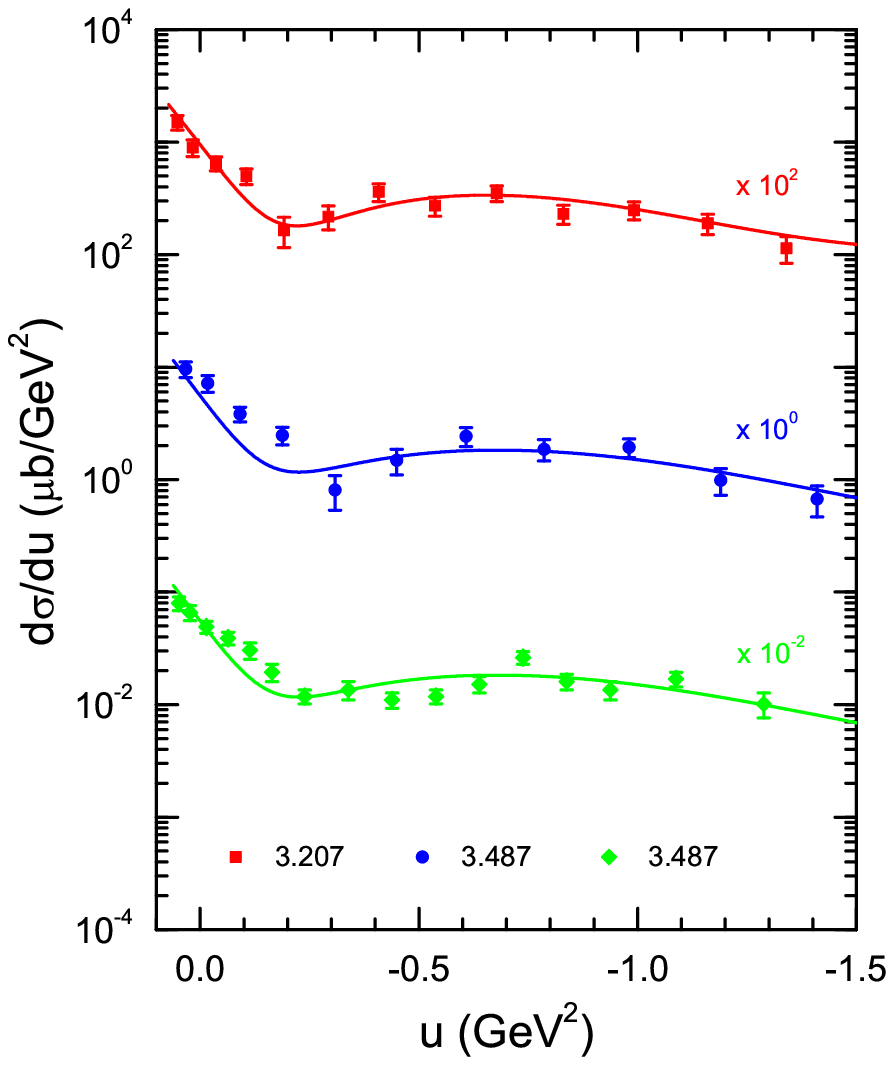}}
\caption{ \label{dsdu_cex} The differential cross section for
$\pi^-p \to \pi^0 n$ backward scattering as a function of the
$u$-channel four-momentum transfer squared shown for different
invariant collision energies indicated in the legend. The references
to the data are given in tab. \ref{cex}. The solid lines are the
results of our model calculation. Both data and calculations were scaled by
the indicated factors.}
\end{figure}

\begin{figure}[t]
\centering \resizebox{0.4\textwidth}{!}{\includegraphics{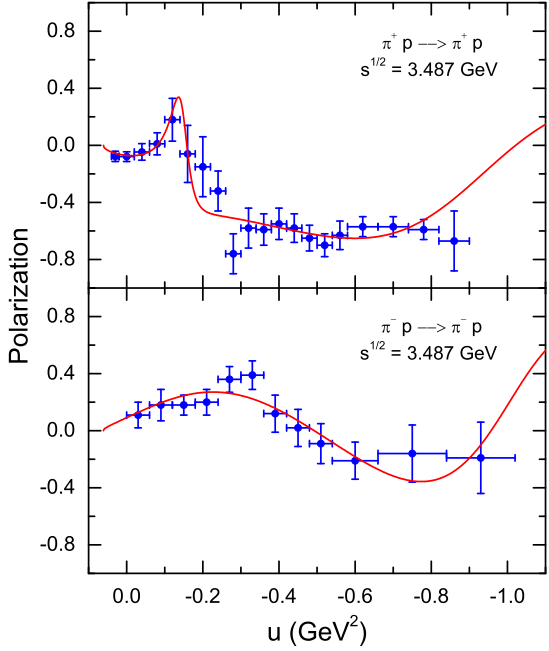}}
\caption{ \label{pol_pi+p_pi-p} The polarization for $\pi^+p \to
\pi^+p$ and $\pi^-p \to \pi^-p$ backward scattering at the pion beam
momentum of 6 GeV ($\sqrt{s}=3.49$ GeV)
as a function of the $u$-channel four-momentum transfer squared. The
references to the data are specified in table~\ref{pol}. The solid
lines are the results of our model calculation. }
\end{figure}

\begin{figure}[t]
\centering
\resizebox{0.4\textwidth}{!}{\includegraphics{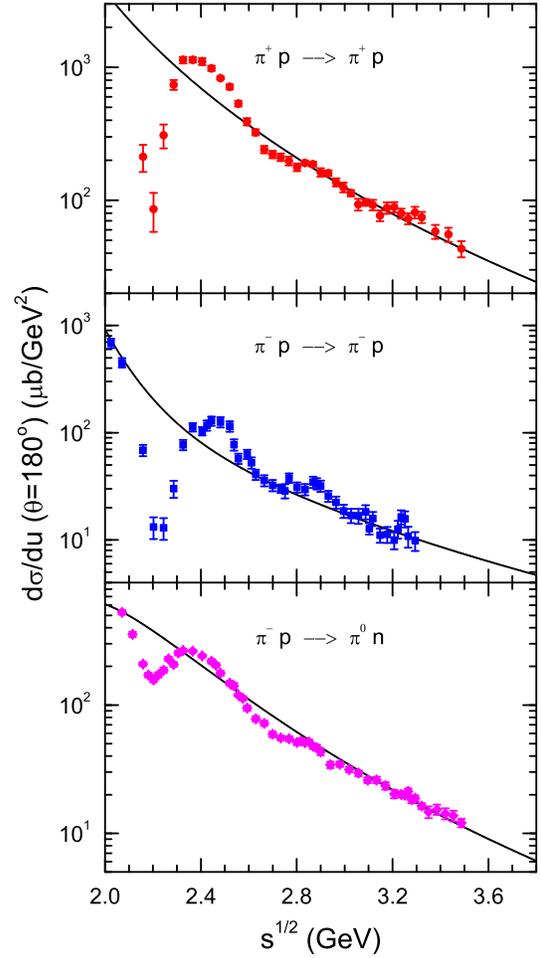} }
\caption{ \label{dsdu180_pi+p_pi-p_cex} Differential cross section
for $\pi^+p \to \pi^+p$, $\pi^-p \to \pi^-p$ and $\pi^-p \to \pi^0n$
reaction at the scattering angle $\theta = 180^\circ$ as a function of the
invariant collision energy. The references to the data are shown in
table \ref{dsdu180}. The solid lines are the results of our model calculation. }
\end{figure}

\begin{figure}[t]
\centering
\resizebox{0.4\textwidth}{!}{\includegraphics{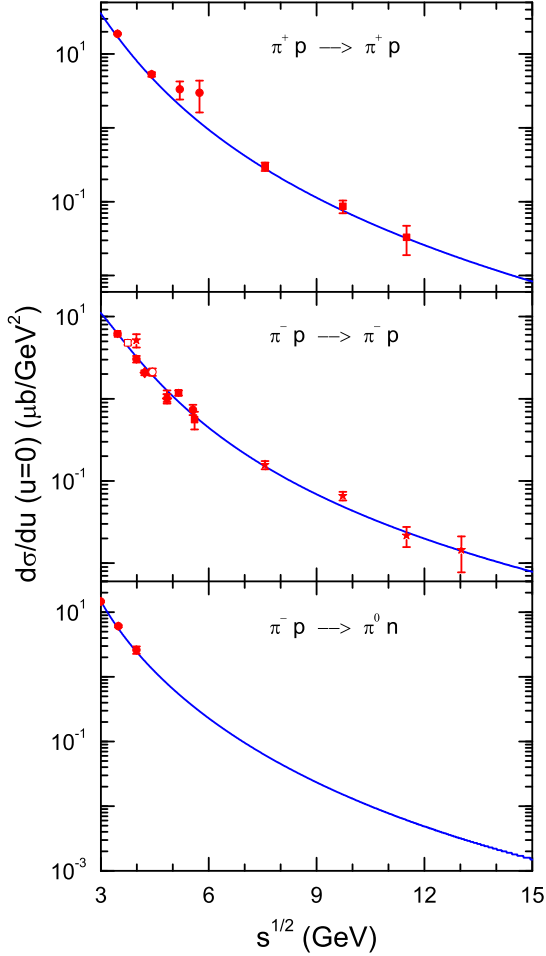}}
\caption{\label{dsduu0_pi+p_pi-p_cex} Differential cross section for
$\pi^+p \to \pi^+p$, $\pi^-p \to \pi^-p$ and $\pi^-p  \to \pi^0n $
scattering at the four-momentum transfer squared $u = 0$ as a function
of the invariant collision energy. The references to the data are shown
in table \ref{dsduu0}. The solid lines show our results. }
\end{figure}

\begin{figure}[t]
\centering
\resizebox{0.4\textwidth}{!}{\includegraphics{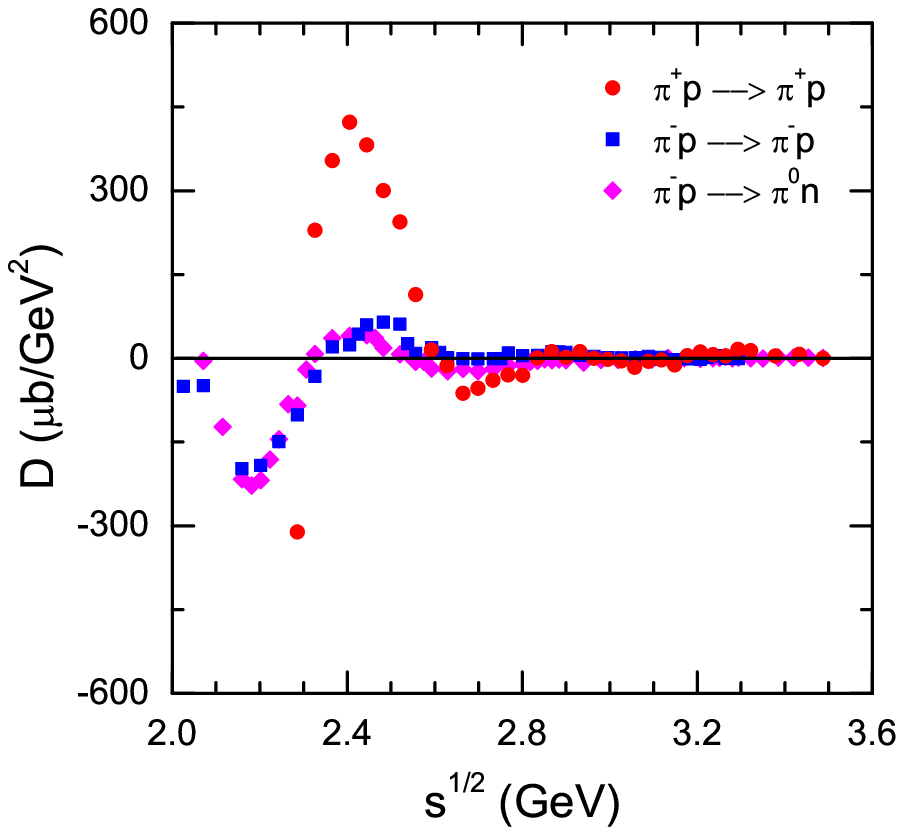}}
\caption{\label{differ}The difference $D$ between the experimental
differential cross sections and the Regge calculation at the scattering
angle $\theta$=180$^0$ as defined by eq. (\ref{difru}), shown as a
function of the invariant collision energy for different reaction
channels.}
\end{figure}

We use almost all data available for the differential cross sections
and polarization asymmetries for backward $\pi N$ scattering 
with invariant collision energy $\sqrt{s} \ge $ 3~GeV, see 
tables \ref{pi-p}-\ref{dsduu0} in the appendix
for a short overview.
For the $\pi^-p \to \pi^0n$ reaction 
(the charge-exchange channel, abbreviated as CEX), 
the data by Boright {\it et al.} \cite{Boright70} 
and Schneider {\it et al.} \cite{Schneider69}
are known to be inconsistent with the experimental results from DeMarzo {\it et al.}
\cite{DeMarzo75} and Chase {\it et al.} \cite{Chase70}. The Regge
model analyses \cite{Berger71,Mir81} done previously included the
data from refs.~\cite{Boright70,Schneider69}. However, in our study
we include the experimental results from refs.~\cite{DeMarzo75,Chase70} 
since these data are more recent and
furthermore for these data the appropriate radiative
corrections have been applied \cite{DeMarzo75,Hoehler83}. For
$\pi^+p \to \pi^+p$ and $\pi^-p \to \pi^-p$ backward scattering,
the data from 
refs.~\cite{Brody66,Bashian74,Ashmore67,Babaev72,Buran76,Babaev77} are not
included in our fitting since they are not consistent with the much more
recent data indicated in the tables.

The problem of discrepancies in the absolute
normalization of the differential cross sections measured in
different experiments is discussed in details in refs.~\cite{Berger71,Mir81}. 
In the present analysis we apply the
procedure proposed in refs.~\cite{Berger71,Mir81} in order to
account for the systematic uncertainties due to the absolute
normalization.

Note that in table~\ref{pol} we indicate references to the data
available at invariant collision energies from 2.35 to 3.49 GeV.
However, only experimental polarization data at energies above
$\sqrt{s}=$ 3~GeV were used in the global fit. 
The other data at low
energies are compared with the Regge calculation in order to
clarify how much they deviate from the expectation based on 
the reaction amplitudes constructed at high energies.

\begin{table}[b]
\begin{center}
\caption{\label{chi2} Summary of the $\chi^2$ for the differential cross
sections and polarization data for $\pi N$ backward scattering. Here
ND denotes the number of data points.}
\renewcommand{\arraystretch}{1.2}
\begin{tabular}{|l|c|c|}
\hline
Observable & ND & $\chi^2\!/$ND  \\
\hline
 $d\sigma/du$ ($\pi^+p$)  & 227 & 2.32   \\
 $d\sigma/du$ ($\pi^-p$)  & 187 & 1.43   \\
 $d\sigma/du$ (CEX)  & ~59 & 1.94  \\
 $P$ ($\pi^+p$)  & ~20 & 0.71  \\
 $P$ ($\pi^-p$)  & ~12 & 0.65  \\
\hline
 Total  & 505 & 1.84  \\
\hline
\end{tabular}
\end{center}
\end{table}

\begin{table}[t]
\begin{center}
\caption{\label{para-u} Parameters of the $N_\alpha$, $N_\gamma$,
$\Delta_\delta$ and $\Delta_\beta$ amplitudes obtained in the global
fit. Note that the slope $\alpha^\prime$ was taken to be the same
for the different amplitudes.}
\renewcommand{\arraystretch}{1.2}
\begin{tabular}{|r|r|r|r|r|}
\hline
Parameters & $N_\alpha$ & $N_\gamma$  & $\Delta_\delta$ & $\Delta_{\beta}$  \\
\hline
 $a$ [GeV$^{-1}$]      & $-60.68$ & $  47.22$   & $ -75.15$ & $ 1419.99$  \\
 $b$ [GeV$^{-3}$]      & $326.52$ & $-215.84$   & $-138.75$ & $ 3052.84$ \\
 $c$ [GeV$^{-2}$]      & $546.40$ & $-101.11$   & $  64.16$ & $-192.64$ \\
 $d$ [GeV$^{-4}$]      & $307.42$ & $-128.04$   & $  86.77$ & $-695.81$ \\
 $\alpha_0$  & $-0.36$ & $ -0.62$   & $   0.03$ & $   -2.65$ \\
\hline
 $\alpha'$ [GeV$^{-2}$]  & \multicolumn{4}{c|}{0.908} \\
\hline
\end{tabular}
\end{center}
\end{table}

Finally, the number of data points (ND) included in the global fit
and the obtained $\chi^2$ for various observables are listed in
table \ref{chi2}. Here we also indicate the $\chi^2/$ND for different
reaction channels. The small $\chi^2$ obtained for the fit to the
polarization data originates from large uncertainties in the 
experimental results. By fitting $505$ data points we get a total
$\chi^2=$~1.84 per data point.

The 21 free parameters of the model are listed in tab.~\ref{para-u}. 
Note that the slope $\alpha^\prime$ was taken to be
the same for the different baryon trajectories. That ensures that
the baryon trajectories are parallel in the plane given by the spin
and mass of baryons.

Figs.~\ref{dsdu_pi+p_c}-\ref{dsdu_pi+p_b} show experimental results
on $\pi^+ p \to \pi^+p$ differential cross sections together with
our calculations. Note that the data indicate a dip near 
$u \simeq -$~0.15~GeV$^2$ where the  $N_\alpha$ amplitude
passes through zero. 
Indeed taking into account the
parameters listed in table~\ref{para-u} it is clear that at $u \simeq
-$~0.15~GeV$^2$ the $N_\alpha$ trajectory becomes $\alpha(u) \simeq -$~1/2
and, therefore, the signature factor given by eq.~(\ref{signature})
is $\zeta_\alpha(u) =$~0. However, the dip in the $\pi^+ p \to \pi^+
p$ differential cross sections is partially filled due to the
contributions from other trajectories that have zeros in the
signature factors at different values of the four-momentum transfer
squared $u$. Also note that at small values of $|u| < $~0.1~GeV$^2$
the data indicate an exponential dependence. 
At $|u| > $~0.4~GeV$^2$ the differential cross sections show
a smooth, almost constant behavior. The dip observed in the  $\pi^+ p
\to \pi^+ p$ differential cross sections allows to conclude that the
$N_\alpha$-trajectory dominates this reaction channel.

In general there is reasonable agreement between the $\pi^+p$
backward scattering data and our Regge calculation. 
There is, however, a disagreement between our
results and the data of ref. \cite{Baker71}
at $\sqrt{s}=$~3.265 GeV in the vicinity of the dip, which signals 
that the Regge approximation starts to break down for low energies.

Figs.~\ref{dsdu_pi-p_a}-\ref{dsdu_pi-p_b} illustrate our
calculation together with experimental results on $\pi^- p \to
\pi^-p$ differential cross sections. Now the data do not have a dip
structure but rather show a smooth $u$-dependence. There is no
$N_\alpha$-trajectory contribution to this reaction channel. The
reaction is entirely governed by the $\Delta$-trajectories.

Taking the parameters from table~\ref{para-u} one might expect
that in the scattering region the first zero of the
$\Delta_\delta$-trajectory is located around $u \simeq -$~1.68~GeV$^2$, 
while the first zero of the $\Delta_\beta$-trajectory is
around $u \simeq -$~2~GeV$^2$. There are no data available at these
four-momentum transfers to clarify the situation. And, moreover, at
these large values of $|u|$ one should expect additional
contributions from the $t$ and $s$ channel exchanges too.

It is interesting that the data shown in fig.~\ref{dsdu_pi-p_a}
indicate some increase of the differential cross section at $|u|
>$~0.8 GeV$^2$, although the accuracy of the experimental results
is not so high. The model calculation does not produce such a trend but
rather suggests that the differential cross sections slightly decrease.

The differential cross section for the  $\pi^- p \to \pi^0n$ charge
exchange reaction is shown in fig.~\ref{dsdu_cex}. The data
indicate a dip around $u \simeq -$~0.15~GeV which originates from the
$N_\alpha$-trajectory.  However, the structure of the dip observed in
the charge exchange reaction differs from the one seen in the
$\pi^+p \to \pi^+p$ differential cross section. Historically this
difference motivated the inclusion of an additional
$N_\gamma$-trajectory in the Regge analysis of pion-nucleon backward
scattering. Our Regge model describes the differential cross
sections available for the $\pi^-p \to \pi^0n$ reaction fairly well.

Concerning the differential cross sections we find that our Regge
calculation reproduces the data available for $\pi^+p \to
\pi^+p$, $\pi^-p \to \pi^-p$ and  $\pi^-p \to \pi^0n$ backward
scattering reasonably well and thus corroborates the finding of the
previous analyses \cite{Barger67,Barger68,Berger71,Gregorich71,Storrow75}
that, in principle, three trajectories, namely $N_\alpha$, $N_\gamma$ and
$\Delta_\delta$ play a significant role in describing the data on
differential cross sections. 
 The most striking feature of the data is the
dip observed for the $\pi^+p \to \pi^+p$ and $\pi^-p \to \pi^0n$
reaction. This dip allows to determine the intercept $\alpha_0$ of the
$N_\alpha$-trajectory. At the same time there is no dip in the
$\pi^-p \to \pi^-p$ backward scattering since there is no
contribution from the $N_\alpha$-trajectory in that reaction
channel.

Fig.~\ref{pol_pi+p_pi-p} shows the polarization for $\pi^+ p \to
\pi^+p$ and $\pi^-p \to \pi^-p$ backward scattering. These data were
collected \cite{Dick72,Dick73} at the pion beam momentum of 6~GeV 
corresponding to an invariant collision energy of $\sqrt{s}=$~3.49~GeV.  
For both reactions the polarization substantially depends on the 
four-momentum transfer squared $u$. Note that only the $\Delta$-trajectories
contribute to the $\pi^-p \to \pi^-p$ backward scattering. Therefore,
within models that only take into account the 
$\Delta_\delta$-trajectory, it is impossible to reproduce the
polarization for the $\pi^-p \to \pi^-p$ reaction.
Previous analyses \cite{Barger67,Barger68,Berger71,Gregorich71,Storrow75}
made many ad hoc assumptions, but did not manage to achieve any consistent fits. 
The present work includes the $\Delta_\beta$-trajectory, which allows to
reproduce the polarization data.

\section{Extrapolation below  3 GeV}

Fig.~\ref{dsdu180_pi+p_pi-p_cex} shows the energy dependence of the
differential cross section for the reactions $\pi^+p \to \pi^+p$, $\pi^-p \to
\pi^-p$ and $\pi^-p \to \pi^0p$ measured at the pion
scattering angle of 180$^0$. The data indicate considerable
structures for center-of-mass energies up to $\simeq$~3~GeV or
even higher. There were intense discussions
\cite{Hoehler83,Lennox75,Ma75,Hendry81} whether these structures
originate from the excitation of high mass baryons. Indeed, these data on
pion scattering at 180$^0$ seem to be up to now the only direct
evidence of the existence of excited baryons with masses above
2.4~GeV.

The lines in fig.~\ref{dsdu180_pi+p_pi-p_cex} show the
results of our calculation extrapolated to low energies. Note that
the data below 3~GeV were not included in our fit. We
observe that the data oscillate around the non-resonant
continuum given by the Regge amplitude. Above $\simeq$ 3 GeV our
calculations approach the experimental data. Although the
differential cross section of the $\pi^-p \to \pi^-p$ reaction
indicates some fluctuations above $\simeq$~3~GeV we consider this as
being due to statistical uncertainties. This point will be illustrated
below. Fig.~\ref{dsduu0_pi+p_pi-p_cex}
displays the data available for the $\pi^+p \to \pi^+p$, $\pi^-p \to
\pi^-p$ and $\pi^-p \to \pi^0n$ reactions at the four-momentum transfer
squared $u =$~0~GeV$^2$ as a function of energy. One sees that the
Regge approach agrees with the data above $\sqrt{s} \simeq $~3~GeV.

Next we investigate whether the fluctuations of the data with
respect to our Regge calculation shown in 
fig.~\ref{dsdu180_pi+p_pi-p_cex} for the energies $2 \le \sqrt{s} \le
3.5$~GeV is of systematic or of statistical nature. For that purpose, we evaluate
the difference $D$ between the experimental differential cross sections
and those predicted by the Regge model at the scattering angle $\theta=$~180$^o$ 
at each $\sqrt{s}$ and for each reaction channel, {\it i. e.}
\begin{eqnarray}
D =\frac{d\sigma^{\rm Exp.}}{du}-\frac{d\sigma^{\rm Regge}}{du}~, 
\label{difru}
\end{eqnarray}
and present the results in fig.~\ref{differ} using a linear scale.

At energies $\sqrt{s} >$~2.8 GeV the data available
for $\pi^+p \to \pi^+p$, $\pi^-p \to \pi^-p$ and $\pi^-p \to \pi^0n$
scattering at $\theta=$~180$^0$ are consistent with the predictions of our
Regge
calculation. Below that energy the data indicate some room for
additional contributions. At least the magnitude of the variations
seem to be larger than the statistical fluctuations of the experimental 
results.
Furthermore, the different $\pi N$ reaction channels do not indicate the same
pattern of differences between the data and the Regge results. The largest 
difference occurs for the reaction
$\pi^+p \to \pi^+p$ at 180$^0$, while the values of $D$ 
obtained for $\pi^-p \to \pi^-p$ and $\pi^-p \to \pi^0n$ are
almost identical. Note that in the $s$-channel only excitations of
$\Delta$-isobars is possible for the $\pi^+p \to \pi^+p$
reaction. The two other reactions allow for the excitation of nucleon as well as 
$\Delta$-resonances in the $s$-channel.

It is interesting to illustrate the arguments of ref.~\cite{Lennox75} 
where similar differences between the Regge predictions
and data for $\pi^+p$ elastic scattering at 180$^0$ were evaluated in
terms of an additional resonance contribution. Note that at
180$^0$ only the spin non-flip amplitude contributes, see eq.~(\ref{flip}).
The amplitude due to the excitation of baryon
resonances in the $s$-channel is given by
\begin{eqnarray}
{\cal M}^{++} = \sum_n \frac{C_n \, X_n \, (J+1/2)}{\epsilon
-i}(-1)^{l}~,
\end{eqnarray}
where the summation is done over the resonances, $C_n$ is a Clebsch-Gordan
coefficient, $X_n$ denotes the resonance elasticity, $J$ the spin of the
resonance, and $l$ the orbital momentum between pion and nucleon.
Here
\begin{eqnarray}
 \epsilon = \frac{M_R^2-s}{M_R\Gamma_R},
\end{eqnarray}
with $M_R$ and $\Gamma_R$ being the resonance mass and width,
respectively. The width was taken as energy-dependent. Furthermore, 
recalling $ P_l(\cos\theta = {-1}) = (-1)^l$,
one sees explicitly that 
the resonance amplitudes interfere with the non-resonant amplitude  
either constructively or destructively according to the parity of the resonance.
In the case of $\pi^+p \to \pi^+p$ elastic scattering
the $s$-channel contributions from the $D_{35}(2350)$ and
$H_{3,11}(2420)$ with $l =$ 2 and $l =$ 5 are the major candidates responsible
for the change of the sign of $D$ seen in fig.~\ref{differ}.


\begin{figure}[t]
\centering
\resizebox{0.45\textwidth}{!}{\includegraphics{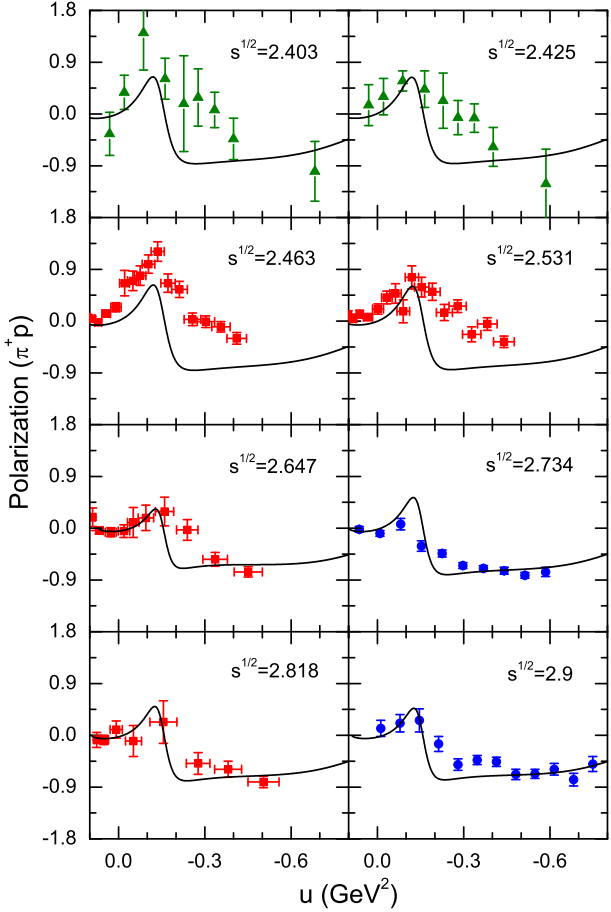}}
\caption{ \label{pol_pi+p} The polarization asymmetry for $\pi^+p
\to \pi^+p$ backward scattering at different invariant collision 
energies as indicated. The references to the
data are specified in tab.~\ref{pol}. The solid lines show the results
of our Regge
calculation with the model parameters listed in table~\ref{para-u}.}
\end{figure}

\begin{figure}[t]
\centering
\resizebox{0.45\textwidth}{!}{\includegraphics{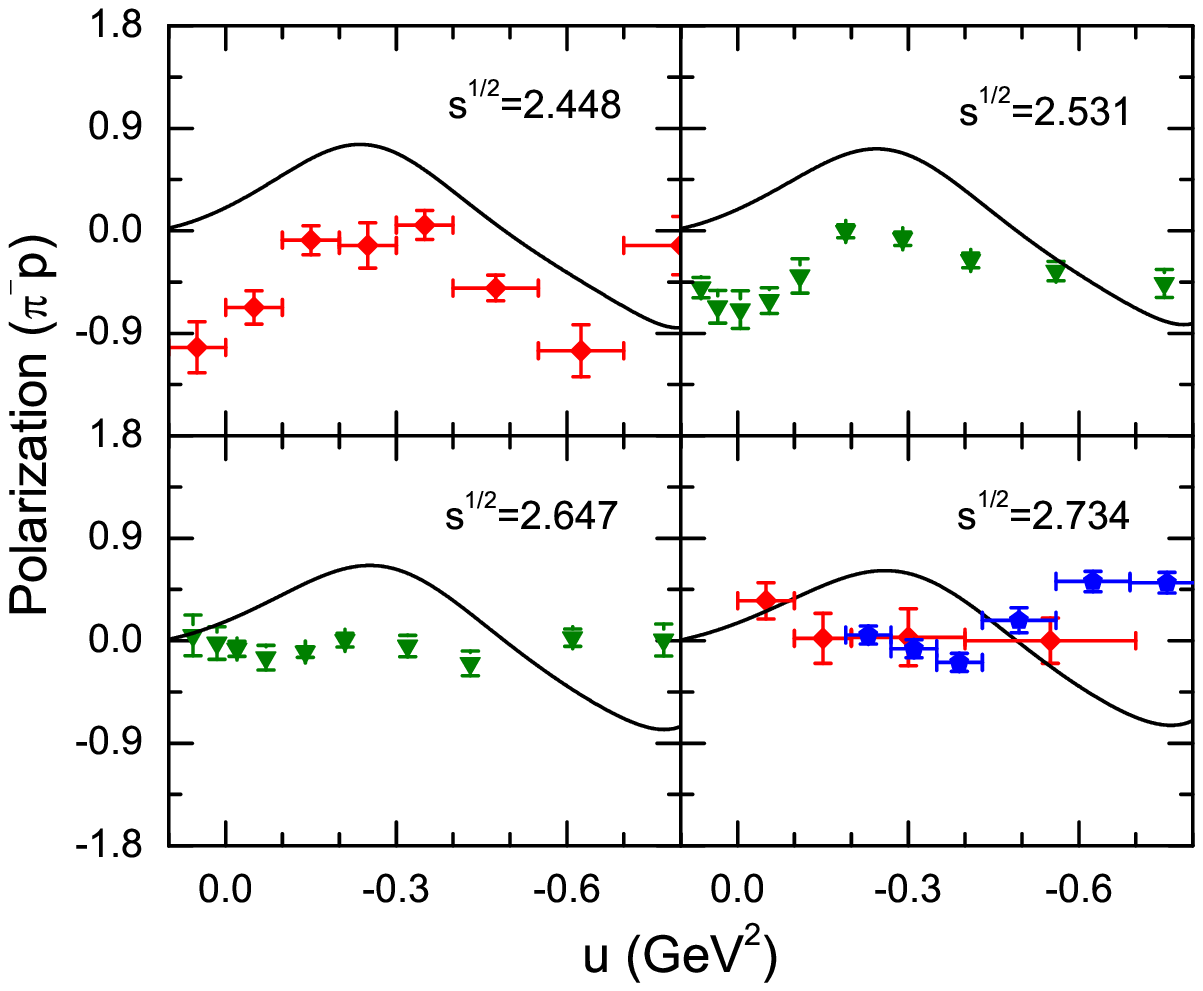}}
\caption{ \label{pol_pi-p} The polarization asymmetry for $\pi^-p
\to \pi^-p$  backward scattering at different
invariant collision energies as indicated. The references to the
data are specified in table~\ref{pol}. The solid lines show the results of 
our Regge
calculation with the model parameters listed in table~\ref{para-u}.}
\end{figure}

\begin{figure}[t]
\centering
\resizebox{0.4\textwidth}{!}{\includegraphics{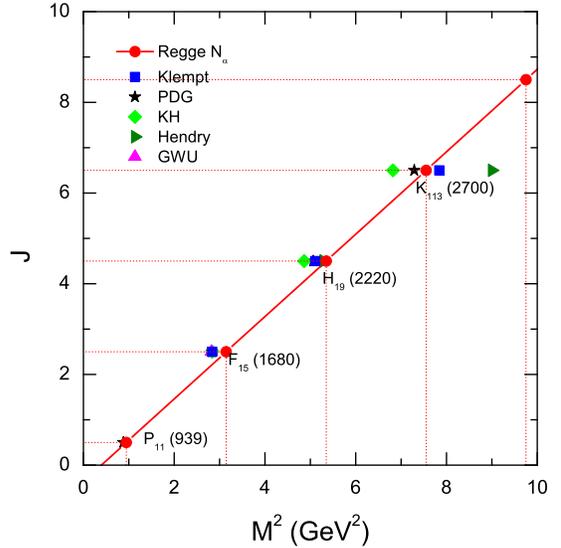}}
\caption{\label{nalpha} Chew-Frautschi plot for the $N_\alpha$
trajectory for baryons with parity $P=+$ 1 and signature ${\cal
S}=+$ 1. The line shows the Regge  trajectory according to table~\ref{para-u}.
 The symbols indicate the
results from other approaches.}
\end{figure}

\begin{figure}[t]
\centering
\resizebox{0.4\textwidth}{!}{\includegraphics{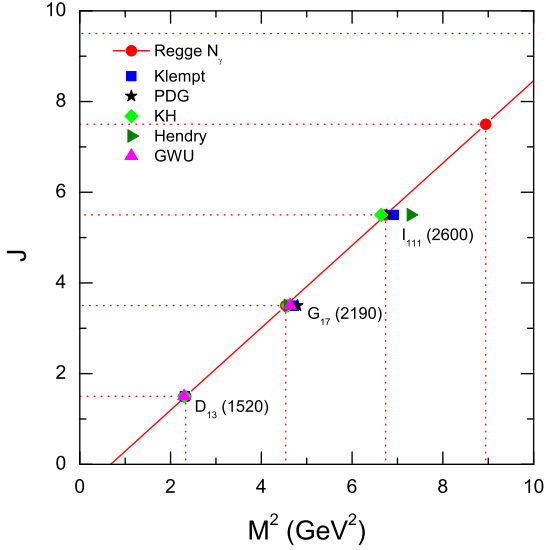}}
\caption{\label{ngamma} $N_\gamma$ trajectory for baryons with
parity $P=-$ 1 and signature ${\cal S}=-$ 1. The line shows the Regge
trajectory according to table~\ref{para-u}.
 The symbols indicate the results from other approaches.}
\end{figure}

\begin{figure}[t]
\centering
\resizebox{0.4\textwidth}{!}{\includegraphics{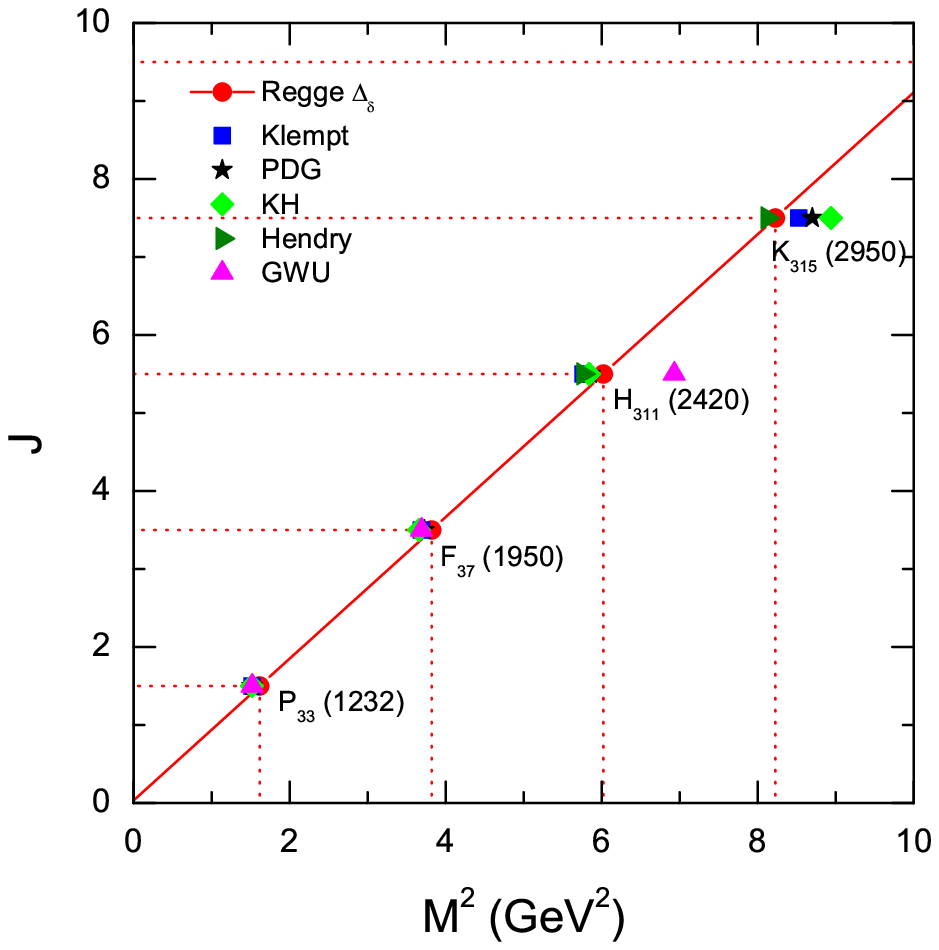}} \caption{
\label{ddelta} $\Delta_\delta$ trajectory for $\Delta$-isobars with
parity $P=+$ 1 and signature ${\cal S}=-$ 1. The line shows the Regge
trajectory, while the symbols indicate the results from other approaches.}
\end{figure}

\begin{figure}[t]
\centering \resizebox{0.4\textwidth}{!}{\includegraphics{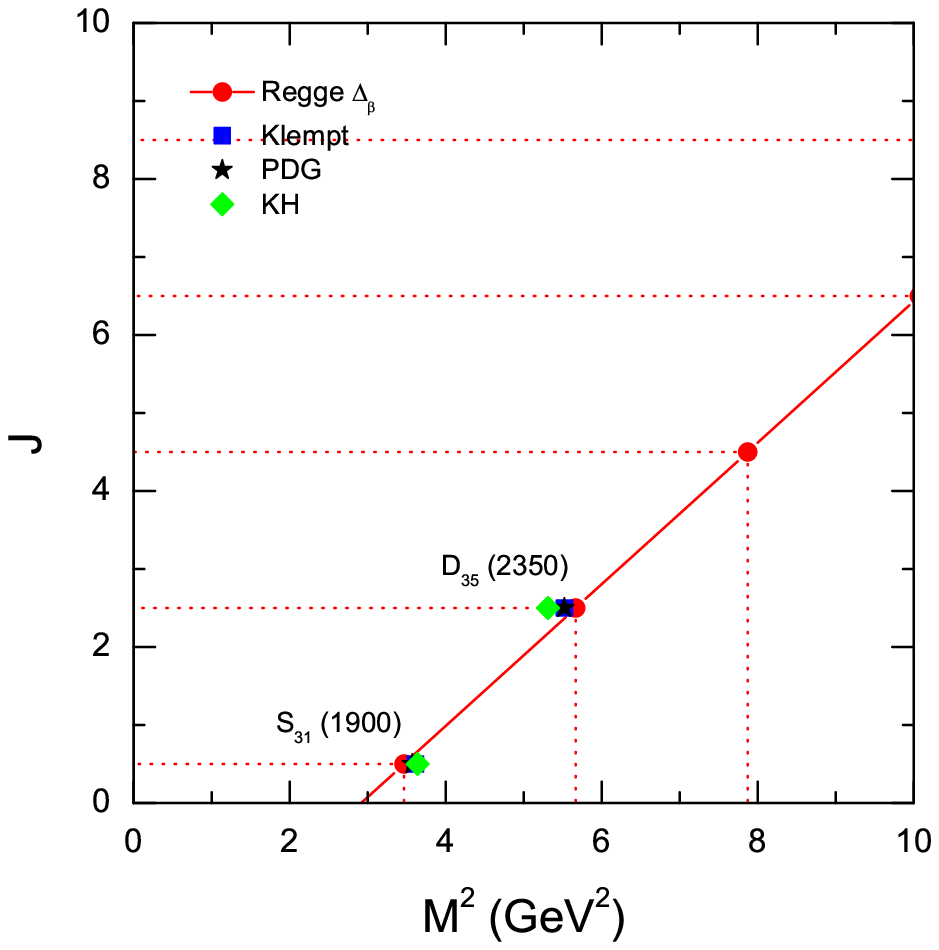}}
\caption{\label{dbeta} $\Delta_\beta$ trajectory for
$\Delta$-isobars with parity $P=-$ 1 and signature ${\cal S}=+$ 1.
The line shows the Regge trajectory, while the symbols indicate the 
results from other approaches.}
\end{figure}

As is illustrated by fig.~\ref{pol_pi+p_pi-p} the Regge model 
reproduces the polarization data for $\pi^+p \to \pi^+p$ and $\pi^-p
\to \pi^-p$ backward scattering at $\sqrt{s}=$ 3.49 GeV. 

In fig.~\ref{pol_pi+p}, we show the 
polarization data for backward scattering
in the reaction $\pi^+p \to \pi^+p$  available in the energy range
$2.4 < \sqrt{s} < 3$ GeV  \cite{Bradamante73,Sherden70}.
For energies above $\sqrt{s} \simeq $~2.73 GeV the Regge results 
reproduce the polarization data reasonably well, but 
within the range $2.46 \le \sqrt{s} \le 2.64$~GeV,
there is  room for additional contributions at $|u| > $~0.1~GeV$^2$.
This finding is consistent with the data on differential cross section 
presented in fig.~\ref{dsdu180_pi+p_pi-p_cex}.

The situation is quite different for the polarization observed in
$\pi^-p \to \pi^-p$ backward scattering \cite{Fukushima80,Birsa76}.
The data \cite{Fukushima80} at $\sqrt{s} \simeq$~2.45~GeV show
a $u$-dependence similar to the one given by the Regge model, but with a
systematic shift to negative values. The data at $\sqrt{s}=$~2.65~GeV 
and $\sqrt{s}=$~2.73~GeV indicate an almost zero
polarization within the experimental uncertainties. This is in line with
the $\pi^-p \to \pi^-p$ data on the differential cross section at
180$^0$ displayed in fig.~\ref{dsdu180_pi+p_pi-p_cex}.

Let us finally come to the baryon trajectories. 
Fig. \ref{nalpha} shows the Chew-Frautschi plot for the $N_\alpha$
trajectory.
The poles of the amplitude of eq.~(\ref{amplitude}) correspond to
baryons with spin~$J$, as is indicated by the dashed lines. The
results of partial wave analyses (PWA) 
from the Karlsruhe-Helsinki (KH) \cite{Hoehler83,Koch80} and
the George Washington University (GWU) \cite{Arndt04} groups 
are included in the figure, too. We also indicate the averaged values
given by the Particle Data Group \cite{PDG} and the most recent systematic analysis
by Klempt and  Richard \cite{Klempt09}. 
The analysis by Hendry~\cite{Hendry78,Hendry81}
is based on an impact parameter approach that 
differs significantly from our result for the $K_{1,13}(2700)$.

Fig.~\ref{ngamma} shows the Chew-Frautschi plot for the $N_\gamma$
trajectory. There are two resonances with a four-star rating by the PDG on this 
trajectory, and even the $I_{1,11}(2600)$ is classified with three stars.

The $\Delta_\delta$- and $\Delta_\beta$-trajectories are shown in
figs.~\ref{ddelta} and \ref{dbeta}, respectively. 
The mass of the $H_{3,11}(2420)$ isobar obtained by the GWU PWA differs
significantly from other results. 
Our analysis suggests a $G_{39}$ resonance with a mass of 2.83~GeV as member
of the $\Delta_{\beta}$ trajectory.

\section{Summary}

We have performed a systematic analysis of the data on differential cross
sections and polarizations available for $\pi^+p \to \pi^+p$, $\pi^-p
\to \pi^-p$ and $\pi^-p \to \pi^0n$ scattering at backward angles.
We started out from a Regge model including the $N_\alpha$, $N_\gamma$,
$\Delta_\delta$ and $\Delta_\beta$ trajectories 
and determined the reaction amplitude at energies $\sqrt{s} >3$~GeV.
In contrast to previous analyses,
the present study reproduces the polarization asymmetriess for both
$\pi^+p$ and $\pi^-p$ backward elastic scattering 
within standard Regge phenomenology. We found that it is not
necessary to resort to non-Regge terms
\cite{Gregorich71,Park74,Mir81,Storrow75,Storrow73} to describe the
polarization, but rather that it is important to include the
$\Delta_\beta$-trajectory which was neglected in
previous analyses.

After the reaction amplitude was fixed at high energies we have inspected
the data on differential cross sections for scattering at $\theta = 180^0$ 
and the polarization asymmetry at energies $2 \le \sqrt{s} \le
3$~GeV. The data available at $\theta = 180^0$ are of special
interest because the $\pi^+p \to \pi^+p$, $\pi^-p \to \pi^-p$, and
$\pi^-p \to \pi^0n$ differential cross sections indicate
considerable structures for center-of-mass energies up to
$\simeq 3$~GeV.
The data fluctuate around the cross sections given
by the Regge calculation. This can be considered as direct evidence of
the excitation of baryons with masses up to approximately 2.8~GeV. 
This point of view is further supported by 
the data on the polarization asymmetry available at these energies. 

\begin{acknowledgement}
This work is partially supported by the Helmholtz Association through funds
provided to the virtual institute ``Spin and strong QCD'' (VH-VI-231), by
the EU Integrated Infrastructure Initiative  HadronPhysics2 Project (WP4 QCDnet) 
and by DFG (SFB/TR 16, ``Subnuclear Structure of Matter''). This work was 
also supported in part by U.S. DOE
Contract No. DE-AC05-06OR23177, under which Jefferson Science Associates,
LLC, operates Jefferson Lab. F.H. is grateful to the support from the
Alexander von Humboldt Foundation during his stay in J\"{u}lich where the
main part of this paper was completed  and the support by COSY FFE grant
No. 41445282 (COSY-058). A.S. acknowledges support by the
JLab grant SURA-06-C0452  and the COSY FFE grant No. 41760632 (COSY-085). 
\end{acknowledgement}

\vfill

\pagebreak

\section*{Appendix:  Data collection}
The appendix summarizes the world data on backward pion-nucleon
scattering.

\begin{table}[htb]
\begin{center}
\caption{\label{pi+p} References to data on $\pi^+p \to \pi^+p$
differential cross sections for backward scattering.}
\renewcommand{\arraystretch}{1.2}
\begin{tabular}{|r|r|r|r|l|c|}
\hline
  $\sqrt{s}$~~ & $p_\pi$~ & $u_{min}$ & $u_{max}$ & Experiment & Ref.  \\
  GeV      & \hspace{-2mm} GeV \hspace{-2mm}  & GeV$^2$ &  GeV$^2$ &
 & \\
\hline
 3.02 & 4.4 & $-$0.04 & 0.07  & Lennox 75 & \cite{Lennox75} \\
 3.06  & 4.5 & $-$0.05 & 0.07  & Lennox 75 & \cite{Lennox75} \\
 3.09  & 4.6 & $-$0.05 & 0.07  & Lennox 75 & \cite{Lennox75} \\
 3.12  & 4.7 & 0.01 & 0.07  & Lennox 75 & \cite{Lennox75} \\
 3.15  & 4.8 & 0.00 & 0.07  & Lennox 75 & \cite{Lennox75} \\
 3.18  & 4.9 & 0.00 & 0.07  & Lennox 75 & \cite{Lennox75} \\
 3.21  & 5.0 & 0.00 & 0.07  & Lennox 75 & \cite{Lennox75} \\
 3.24  & 5.1 & $-$0.00 & 0.07  & Lennox 75 & \cite{Lennox75} \\
 3.26  & 5.2 & $-$0.00 & 0.06  & Lennox 75 & \cite{Lennox75} \\
 3.26  & 5.2 & $-$0.82 & 0.01  & Baker 71 & \cite{Baker71} \\
 3.29  & 5.3 & $-$0.01 & 0.06  & Lennox 75 & \cite{Lennox75} \\
 3.32  & 5.4 & $-$0.01 & 0.06  & Lennox 75 & \cite{Lennox75} \\
 3.38  & 5.6 & $-$0.01 & 0.06  & Lennox 75 & \cite{Lennox75} \\
 3.43  & 5.8 & $-$0.01 & 0.06  & Lennox 75 & \cite{Lennox75} \\
 3.46  & 5.9 & $-$0.87 & 0.06  & Owen 69  & \cite{Owen69} \\
 3.49  & 6.0 & $-$0.02 & 0.06  & Lennox 75 & \cite{Lennox75} \\
 3.75  & 7.0 & $-$1.15 & 0.01  & Baker 71 & \cite{Baker71} \\
 3.99  & 8.0 & $-$0.06 & 0.04  & Frisken 65 & \cite{Frisken65} \\
 3.99  & 8.0 & $-$0.06 & 0.04  & Orear 66 & \cite{Orear66} \\
 4.40  & 9.8 & $-$2.29 & 0.03  & Owen 69 & \cite{Owen69}  \\
 4.43  & 10.0 & $-$17.45 & $-$0.62  & Baglin 75 & \cite{Baglin75} \\
 5.16  & 13.7 & $-$2.82 & 0.01 & Owen 69 & \cite{Owen69}  \\
 5.74  & 17.1 & $-$0.01 & 0.01 & Owen 69 & \cite{Owen69}  \\
 7.56  & 30.0 & $-$0.55 & 0.00 & Baker 83 & \cite{Baker83} \\
 9.73  & 50.0 & $-$0.50 & 0.00 & Baker 83 & \cite{Baker83} \\
11.50  & 70.0 & $-$0.05 & $-$0.01 & Baker 83 & \cite{Baker83} \\
\hline
\end{tabular}
\end{center}
\end{table}

\begin{table}[htb]
\begin{center}
\caption{\label{cex} References to data on $\pi^-p \to \pi^0n$
differential cross sections for backward scattering.}
\renewcommand{\arraystretch}{1.2}
\begin{tabular}{|r|r|r|r|l|c|}
\hline
  $\sqrt{s}$~~ & $p_\pi$~ & $u_{min}$ & $u_{max}$ & Experiment & Ref.  \\
  GeV      &  GeV  & GeV$^2$ &  GeV$^2$ &   &   \\
\hline
 3.21  & 5.0 & $-$1.91 & 0.05  & Chase 70 & \cite{Chase70} \\
 3.49  & 6.0 & $-$2.12 & 0.03  & Chase 70 & \cite{Chase70} \\
 3.49  & 6.0 & $-$1.29 & 0.05  & DeMarzo 75 & \cite{DeMarzo75} \\
\hline
\end{tabular}
\end{center}
\end{table}

\begin{table}[ht]
\begin{center}
\caption{\label{pi-p} References to data on $\pi^-p \to \pi^-p$
differential cross sections for backward scattering.}
\renewcommand{\arraystretch}{1.2}
\begin{tabular}{|r|r|r|r|l|c|}
\hline
  $\sqrt{s}$~~ & $p_\pi$~ & $u_{min}$ & $u_{max}$ & Experiment & Ref.  \\
  GeV      & \hspace{-2mm} GeV \hspace{-2mm}  & GeV$^2$ &  GeV$^2$ &   &
 \\
\hline
 3.46  & 5.9 & $-$0.89 & 0.06  & Owen 69 & \cite{Owen69} \\
 3.75  & 7.0 & $-$0.60 & 0.02  & Baker 71 & \cite{Baker71} \\
 3.99  & 8.0 & $-$0.39 & 0.01  & Anderson 68 & \cite{Anderson68} \\
 3.99  & 8.0 & $-$0.06 & 0.04  & Frisken 65 & \cite{Frisken65} \\
 3.99  & 8.0 & $-$0.06 & 0.05  & Orear 66 & \cite{Orear66} \\
 4.22  & 9.0 & $-$1.40 & $-$0.02  & Jacholkowski 77 &
\cite{Jacholkowski77} \\
 4.40  & 9.8 & $-$2.39 & 0.03  & Owen 69 & \cite{Owen69} \\
 4.44  & 10.0 & $-$1.15 & $-$0.07  & Ghidini 82 & \cite{Ghidini82}  \\
 4.84  & 12.0 & $-$1.46 & $-$0.03  & Jacholkowski 77 &
\cite{Jacholkowski77} \\
 5.16  & 13.7 & $-$0.18 & 0.02  & Owen 69 & \cite{Owen69} \\
 5.56  & 16.0 & $-$0.73 & $-$0.09  & Anderson 68 & \cite{Anderson68} \\
 5.60  & 16.3 & $-$0.17 & 0.00  & Owen 69 & \cite{Owen69} \\
 7.56  & 30.0 & $-$0.38 & 0.00  & Baker 83 & \cite{Baker83}  \\
 9.73  & 50.0 & $-$0.44 & $-$0.00  & Baker 83 & \cite{Baker83} \\
11.50  & 70.0 & $-$0.25 & $-$0.01  & Baker 83  & \cite{Baker83} \\
13.00  & 90.0 & $-$0.25 & $-$0.01  & Baker 83  & \cite{Baker83} \\
\hline
\end{tabular}
\end{center}
\end{table}

\begin{table}[htb]
\begin{center}
\caption{\label{pol} References to polarization asymmetry $P$
data for $\pi N$ backward scattering. Note that only data for
$\sqrt{s} > $ 3 GeV were included in our fit.}
\renewcommand{\arraystretch}{1.2}
\begin{tabular}{|c|c|r|r|r|l|c|}
\hline
 & $\sqrt{s}$ & $p_\pi$ & $u_{min}$ & $u_{max}$ & Experiment & Ref.  \\
          &  GeV & \hspace{-2mm} GeV \hspace{-2mm}  & GeV$^2$ & GeV$^2$ &
&  \\
\hline
 $\pi^+p$ & 2.40 & 2.59 & $-$1.40 & 0.03  & Martin 75 & \cite{Martin75} \\
 $\pi^+p$ & 2.43 & 2.65 & $-$1.31 & 0.03  & Martin 75 & \cite{Martin75} \\
 $\pi^-p$ & 2.45 & 2.71 & $-$1.00 & 0.05  & Fukushi. 80 &
\cite{Fukushima80} \\
 $\pi^+p$ & 2.46 & 2.75 & $-$0.41 & 0.11  & Sherden 70 & \cite{Sherden70}
\\
 $\pi^+p$ & 2.53 & 2.93 & $-$0.44 & 0.10  & Sherden 70 & \cite{Sherden70}
\\
 $\pi^-p$ & 2.53 & 2.93 & $-$0.75 & 0.06  & Sherden 70 & \cite{Sherden70}
\\
 $\pi^+p$ & 2.65 & 3.25 & $-$0.45 & 0.09  & Sherden 70 & \cite{Sherden70}
\\
 $\pi^-p$ & 2.65 & 3.25 & $-$0.77 & 0.06 & Sherden 70 & \cite{Sherden70} \\
 $\pi^+p$ & 2.73 & 3.50 & $-$0.59 & 0.06  & Bradam. 73 &
\cite{Bradamante73} \\
 $\pi^-p$ & 2.73 & 3.50 & $-$0.55 & $-$0.05 & Fukushi. 80 &
\cite{Fukushima80} \\
 $\pi^-p$ & 2.73 & 3.50 & $-$0.89 & $-$0.23 & Birsa 76 & \cite{Birsa76} \\
 $\pi^+p$ & 2.82 & 3.75 & $-$0.50 & 0.08  & Sherden 70 & \cite{Sherden70}
\\
 $\pi^+p$ & 2.90 & 4.00 & $-$0.75 & $-$0.01  & Bradam. 73 &
\cite{Bradamante73} \\
 $\pi^+p$ & 3.49 & 6.00 & $-$0.86 & 0.03 & Dick 72 & \cite{Dick72} \\
 $\pi^-p$ & 3.49 & 6.00 & $-$0.93 & $-$0.03 & Dick 73 & \cite{Dick73} \\
\hline
\end{tabular}
\end{center}
\end{table}

\begin{table}[htb]
\begin{center}
\caption{\label{dsdu180} References to differential cross
section data for $\pi N$ scattering at $\theta=180^\circ$. CEX
denotes the $\pi^-p \to \pi^0n$ charge exchange reaction.}
\renewcommand{\arraystretch}{1.2}
\begin{tabular}{|c|r|r|l|c|}
\hline
 Reaction & ${\sqrt{s}}_{min}$ & ${\sqrt{s}}_{max}$ & Experiment & Ref.  \\
          & GeV              & GeV              &    &   \\
\hline
 $\pi^+p$ &  3.03 & 3.49 & Lennox 75 & \cite{Lennox75} \\
 $\pi^-p$ &  3.03 & 3.29 & Kormanyos 66 & \cite{Kormanyos66} \\
 CEX &  3.02 & 3.49 & Kistiakowsky 72 & \cite{Kistiakowsky72} \\
\hline
\end{tabular}
\end{center}
\end{table}

\begin{table}[htb]
\begin{center}
\caption{\label{dsduu0} References to differential cross section
data for $\pi N$ scattering at $u =$ 0 GeV$^2$.}
\renewcommand{\arraystretch}{1.2}
\begin{tabular}{|c|r|r|l|c|}
\hline
 Reaction & ${\sqrt{s}}_{min}$ & ${\sqrt{s}}_{max}$ & Experiment & Ref.  \\
          & GeV              & GeV              &    &   \\
\hline
 $\pi^+p$ &  3.46 & 5.74 & Owen 69 & \cite{Owen69} \\
 $\pi^+p$ &  7.56 & 11.50 & Baker 83 & \cite{Baker83} \\
 $\pi^-p$ &  3.75 & 3.75 & Baker 71 & \cite{Baker71} \\
 $\pi^-p$ &  3.46 & 5.60 & Owen 69 & \cite{Owen69} \\
 $\pi^-p$ &  7.56 & 13.03 & Baker 83 & \cite{Baker83} \\
 $\pi^-p$ &  3.99 & 5.56 & Anderson 68 & \cite{Anderson68} \\
 $\pi^-p$ &  4.22 & 4.84 & Jacholkowski 77 & \cite{Jacholkowski77} \\
 $\pi^-p$ &  4.44 & 4.43 & Ghidini 82 & \cite{Ghidini82} \\
 $\pi^-p$ &  3.99 & 4.84 & Armstrong 87 & \cite{Armstrong87} \\
 CEX &  3.49 & 3.99 & DeMarzo 75 & \cite{DeMarzo75} \\
\hline
\end{tabular}
\end{center}
\end{table}

\end{document}